\documentclass[RNAAS]{aastex631}

\begin{document}

\title{\texttt{FluxCT}: A Web Tool for Identifying Contaminating Flux in \textit{Kepler} and \textit{TESS} Target Pixel Files}

\correspondingauthor{Jessica Schonhut-Stasik}
\email{jessica.s.stasik@vanderbilt.edu}

\author[0000-0002-1043-8853]{Jessica Schonhut-Stasik}
\affiliation{Department of Physics and Astronomy, Vanderbilt University, Nashville, TN 37235, USA}
\altaffiliation{NISE Fellow, Frist Center for Autism + Innovation, Nashville, TN 37212, USA}

\author[0000-0002-3481-9052]{Keivan Stassun}
\affiliation{Department of Physics and Astronomy, Vanderbilt University, Nashville, TN 37235, USA}




\begin{abstract}

\textit{Accepted by Research Notes of the American Astronomical Society --- February 2023.}

We announce \texttt{FluxCT}, a web tool for identifying contaminating flux in \textit{Kepler} and \textit{TESS} target pixel files (TPFs) due to secondary visual sources. We demonstrate the usage of this tool and discuss the benefits of this tool over a simple \textit{Gaia} radius search. \texttt{FluxCT} focuses on clarity and simplicity, where the only input needed from the user is a KIC or TIC ID. 
By more appropriately accounting for the actual shape of the photometric pixel apertures, \texttt{FluxCT} can produce much more accurate estimates of contaminating flux than simple radial cone searches.

\end{abstract}



\section{Introduction} \label{sec:intro} 
Contaminating flux from unexpected sources is a known issue in both \textit{Kepler} \citep{Borucki2010} and \textit{TESS} \citep{Ricker2015} light curves. Excess flux occurs due to a combination of pixel size (4\farcs{} and 21\farcs{}, respectively) and flux integration --- taking the sum of observed flux in all pixels comprising the Target's Pixel Aperture (TPA) to ensure collection of all light from the target. Data for each source is collected in Target Pixel Files (TPFs) and overlaid with a TPA. TPAs vary in size and shape for each source and are approximately 20\farcs{} ($\sim$5 pixels) in the x direction and 40\farcs{} ($\sim$10 pixels) in the y direction\footnote{Based on the average TPA found in our test of 147 \textit{Kepler} sources. However, TPA size varies with brightness, with brighter stars having larger TPAs}. Subsequently, a light curve is created by plotting the integrated flux measurements.  

Measurements determined using light curves containing excess flux may result in inaccurate conclusions. Some documented examples include false-positive planet transit signals caused by eclipsing binaries \citep{Ziegler2016}, the underestimation of planet radii \citep{Ziegler2018}, the possible dilution of stellar oscillations \citep{SchonhutStasik2017, SchonhutStasik2020}, and observed stellar flares attributed to an incorrect source at a rate of 6.7$\pm$0.4\% for \textit{Kepler} short cadence data \citep{Jackman2021}. 

\begin{figure}
\includegraphics[scale=0.4]{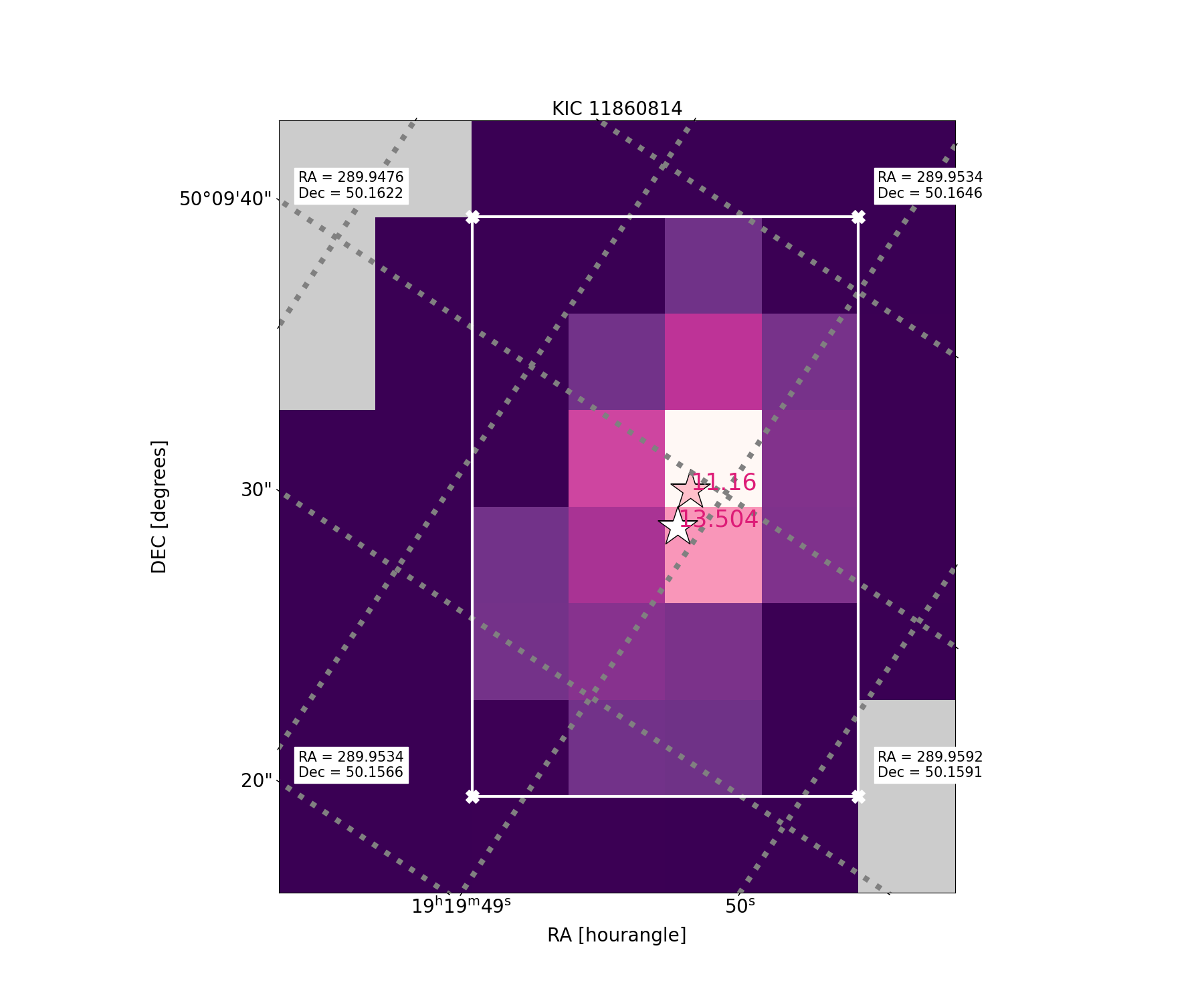}
\caption{Returned plot for KIC 11860814. The pink star shows the target source in RA (hour angle) and Dec (degrees); grey dotted lines represent on-sky coordinates. Each colored square depicts a pixel. Grey pixels constitute areas without data, dark purple represents pixels not included in the TPA, and Lighter shades of purple through white correspond to the source's TPA, scaled toward white as flux values increase. White lines around the TPA show the smallest possible polygon, with RA and Dec values in each corner. KIC 11860814 has one additional source within the TPA, with a target magnitude of 11.16 and 13.50 for the accompanying source, giving a magnitude difference of $\sim$2. There is a flux ratio ($F_2/F_1$) of $\sim$0.1, resulting in a total flux contamination of 10\%. The secondary source has a RUWE of 1.7, suggesting an unresolved double system.}
\label{fig:kic_plot2}
\end{figure}

\section{Web Tool} \label{sec:webtool} 

Using the \texttt{FluxCT} web tool requires only an internet connection and browser access. We built \texttt{FluxCT} by combining \texttt{Python} scripts into a Python Anywhere\footnote{\url{https://www.pythonanywhere.com}} framework, allowing it to run through a website. The user navigates to the site and enters the KIC or TIC ID they wish to search; \texttt{FluxCT} then produces the plots and data to the user's browser.

Once provided an ID number, \texttt{FluxCT} pulls the source's TPF and TPA using the \texttt{Python} module \texttt{lightkurve} \citep{Lightkurve2018}. Then, using the TPA, \texttt{FluxCT} creates a unique search polygon containing the TPA, with as few surrounding pixels as possible. Next, the script searches this area using \texttt{astroquery} \citep{Ginsburg2019} (which utilizes \texttt{astropy} \citep{Astropy2022}) to mine the \textit{Gaia} EDR3 database \citep{Gaia2021} for any potentially contaminating sources in the search polygon.

\texttt{FluxCT} finds potential contaminating sources using \textit{Gaia}, outputting results and a plot; Figure \ref{fig:kic_plot2} shows an example plot. Returned data include all source magnitudes, the magnitude differences between each accompanying source and the target, the ratio of flux between each accompanying source and the target, and the total percentage of flux in the system due to contaminants. For flux, we use \textit{Gaia} G-band values as the closest approximation to the \textit{Kepler} passband. \texttt{FluxCT} also returns the Renormalized Unit Weight Error (RUWE); often used as a marker for close binarity \citep{Lindegren2018, Stassun2021}. The use of RUWE supports completeness, as \textit{Gaia} can only resolve stars down to 1\farcs{} \citep{Ziegler2018}.

\section{Use Benefits} \label{sec:use_case} 

When searching for contaminating sources, it is crucial to consider the shape and size of the search area. For both \textit{Kepler} and \textit{TESS}, the TPA encapsulates many pixels, not just the pixel with the target source. Furthermore, the target source may not fall in a pixel central to the TPA, which is usually asymmetric. Therefore, simple radial searches can prove inadequate for finding the correct amount of contaminating flux. 

In the case of \textit{Kepler}, standard practice is to check for contaminating sources using a 4\farcs{} radial search. The dominant issue is that small radial searches miss possible contaminants. We tested this using a sample of 147 \textit{Kepler} stars with observed oscillations currently undergoing analysis to consider dilution of their oscillation amplitudes (Schonhut-Stasik in prep.). A 4\farcs{} \textit{Gaia} radial cone search found 23 contaminant sources in this sample. In contrast, \texttt{FluxCT} found 107 stars within its unique search polygons, representing a much more complete estimate of the true contaminating flux.  

Sometimes, in an attempt to find all possible contaminating sources, a much larger radial search is used. This more considerable size risks exceeding TPA boundaries and can return many outside stars that would not contribute excess flux. For example, we performed a \textit{Gaia} radial search of 20\farcs{}, where we found 671 sources within 20\farcs{} of our 147 sources, meaning the cone search found 564 stars not identified in the above \texttt{FluxCT} determination---a vast overestimate. These extensive radial searches can be even more problematic for targets that fall closer to the Galactic Plane or in areas of high stellar density.

To be sure, \texttt{FluxCT} is not perfect and will slightly overestimate the number of contaminating sources when the \texttt{FluxCT} rectangular polygon is somewhat larger than the complex TPA. For the test sample, a manual investigation found seven stars within the \texttt{FluxCT} polygon, but outside the true TPA, an overestimate of 6.5\%.

\section{Accessibility} \label{sec:access} 
\texttt{FluxCT} is currently available as a web tool that allows the search of single targets (\url{http://jstasik.pythonanywhere.com}). Creating an accessible web tool is motivated by the desire to encourage broader access to space telescope data for students. The ease of use allows students as early as high school to explore the data without needing advanced knowledge of \texttt{Python}. A full version that can search multiple sources is available at the GitHub \url{https://github.com/Jesstella/FluxCT}. The corresponding author welcomes any suggestions for updates. More example plots and supplemental material exist at \url{https://www.jessicastasik.com/flux-contamination-tool}. A frozen version of the code is available on Zenodo \cite{Fluxct2023}.

\begin{acknowledgments}
This paper includes Kepler mission data obtained from the MAST data archive at the Space Telescope Science Institute (STScI). 
We make use of data from the European Space Agency (ESA) mission \textit{Gaia} (\url{https://www.cosmos.esa.int/gaia}), processed by the \textit{Gaia} Data Processing and Analysis Consortium (DPAC,
\url{https://www.cosmos.esa.int/web/gaia/dpac/consortium}).
\end{acknowledgments}

%

\vspace{5mm}






\bibliography{sample631}{}
\bibliographystyle{aasjournal}



\end{document}